\documentstyle[preprint,prb,aps]{revtex}

\begin{document}

\draft

\title{CVM studies on the atomic ordering in complex perovskite alloys}

\author{Zhi-Rong Liu}
\address{Department of Modern Applied Physics, Tsinghua University, 
Beijing 100084, People's Republic of China}

\author{Jian-She Liu}
\address{Center for Advanced Study, Tsinghua University, 
Beijing 100084, People's Republic of China}

\author{Bing-Lin Gu}
\address{Department of Modern Applied Physics, Tsinghua University, 
Beijing 100084, People's Republic of China}

\author{Xiao-Wen Zhang}
\address{Department of Materials Science and Engineering, 
Tsinghua University, Beijing 100084, People's Republic of China}

\maketitle

\begin{abstract}
The atomic ordering in complex perovskite alloys is investigated by the
cluster variation method (CVM). For the 1/3\{111\}-type ordered structure, 
the order-disorder phase transition is the first order, and the order 
parameter of the 1:2 complex perovskite reaches its maximum near x=0.25. 
For the 1/2\{111\}-type ordered structure, the 
ordering transition is the second order. Phase diagrams for both 
ordered structures are obtained. The order-disorder line obeys the 
linear law.

\end{abstract}


\vspace{2mm}


\section{Introduction}

  Complex perovskite alloys draw immense attention because of their
exceptional dielectric and piezoelectric properties, which have great actual
or potential uses.\cite{1,2} Examples are the so-called ``super $Q$''
mixed-metal perovskites with low losses\cite{3,4}, such as Ba(Zn$_{1/3}$Ta$_{2/3}$)O$_3$ 
(BZT) and Ba(Mg$_{1/3}$Nb$_{2/3}$)O$_3$-BaZrO$_3$ (BMN-BZ), and the relaxor 
ferroelectrics (relaxors) with extraordinarily high values of the 
relative permittivity and the piezoelectric constants\cite{5,6}, such as 
Pb(Sc$_{1/2}$Ta$_{1/2}$)O$_3$-PbTiO$_3$ (PST-PT) and 
Pb(Mg$_{1/3}$Nb$_{2/3}$)O$_3$
-PbTi$O_3$ (PMN-PT). An important aspect in complex perovskite alloys is the 
compositional atomic ordering, which is closely related to the desired 
properties. For example, it was indicated that the microwave-loss 
properties of 1:2 perovskite ceramics are very sensitive to the B-site 
cation ordering.\cite{3} In relaxors, it is widely accepted that the 
presence of the nanoscale ordered 
microregions is responsible for the relaxor behaviors.\cite{7,8}
Some investigations have been done on the atomic ordering of 
complex perovskites. Considering the electrostatic interactions 
between the nearest B-site ions, Zhang {\it et al.} presented an available 
quantitative criterion of the order-disorder transition and 
successfully explained many experimental results in A(B$'_{1/2}$B$''_{1/2}$)O$_3$.
\cite{9} Recently, Bellaiche and Vanderbilt used the electrostatic model to 
satisfactorily reproduce the various types of ordered structures in 
BMN-BZ and other complex perovskites.
\cite{10} In addition, the first-principle calculations were conducted to 
describe the ordering in some cases.\cite{11,12,13}

  It has been revealed\cite{14,15} that the Fourier transforms of 
the atomic interactions play important roles in the atomic ordering. 
In this paper, we adopt the parameters in the reciprocal space of 
atomic interactions and conduct a 
cluster variation method (CVM)\cite{16} calculation to investigate 
the order-disorder transitions in complex perovskites.

\section{Method}

  When considering the atomic ordering in complex perovskites, we 
can model $(1-x)$A(B$'_c$B$''_{1-c}$)O$_3$-$x$AB$'''$O$_3$ as a three-component 
system ($1-x$)(B$'_c$B$''_{1-c}$)$\cdot x$B$'''$ on a simple cubic Bravais 
lattice. $c$=1/3 for 1:2 perovskites such as BMN-BZ, and $c$=1/2 for 
1:1 perovskites such as PST-PT.

  The driving mechanism responsible for the ordering is mainly the 
electrostatic interaction.\cite{9,10} Introduce $\sigma({\bf R})=$
-1, 0, or +1 in the case of $c=1/2$ if ${\bf R}$ is occupied by B$'$, 
B$'''$ or B$''$. In the case of $c=1/3$, introduce $\sigma({\bf R})=$
-2, 0, or +1 for B$'$, B$'''$ or B$''$. The charge fluctuation of 
ions at {\bf R}, $\Delta q({\bf R})$, is proportional to 
$\sigma({\bf R})$.\cite{10} The electrostatic 
energy between a pair of cations at {\bf R} and {\bf R}$'$ is 
proportional to $\Delta q({\bf R})\Delta q({\bf R}')$ or 
$\sigma({\bf R})\sigma'({\bf R}')$, i.e.,

\begin{equation}
E_{elec}(\sigma,\sigma',{\bf R},{\bf R}')
=W({\bf R-R'})\sigma({\bf R})\sigma'({\bf R}'),
\end{equation}
where $W({\bf R-R}')$ is the effective-interaction energy. In a 
purely Coulomb picture\cite{10}, $W({\bf R})\propto 1/R$. In the 
nearest-neighbor Coulomb picture\cite{9}, $W({\bf R})$ is equal to 
zero unless {\bf R} is the nearest neighbor. Different from the two
pictures, we do not make any assumption on the form of $W({\bf R})$ here. 
The energy of the system with a certain configuration 
$\{\sigma({\bf R})\}$ can be expressed as

\begin{equation}
E\left(\{\sigma({\bf R})\}\right)
=E_0+\frac{1}{2}\sum\limits_{{\bf R,R'}}
W({\bf R-R'})\sigma({\bf R})\sigma'({\bf R'}),
\end{equation}
where $E_0$ is a constant independent on 
$\{\sigma({\bf R})\}$.The average value of the energy is 
($E_0$ is neglected)

\begin{eqnarray}
E &=&\sum\limits_{\{\sigma({\bf R})\}}E(\{\sigma({\bf R})\})
P(\{\sigma({\bf R})\}) \nonumber  \\
  &=&\frac{1}{2}\sum\limits_{{\bf R,R'},\sigma,\sigma'}
     W({\bf R-R'})\sigma({\bf R})\sigma'({\bf R'})
              X_{\sigma\sigma'}({\bf R,R'}),
\end{eqnarray}
where $P(\{\sigma({\bf R})\})$ is the statistic probability of the configuration 
$\{\sigma({\bf R})\}$, and $X_{\sigma\sigma'}({\bf R,R'})$ is 
the pair occupation probability.

  When the atomic interaction is long range, Eq. (3) is difficult to 
handle in CVM because a 
large cluster should be used to contain the long-range interaction. 
Therefore, some approximation should be made. 

  Expand $X_{\sigma\sigma'}({\bf R,R'})$ in a Fourier transform:
\begin{equation}
X_{\sigma\sigma'}({\bf R,R'})
=\sum\limits_{\bf k,k'}Q_{\sigma\sigma'}({\bf k,k'})
    \exp[i({\bf k\cdot R+k'\cdot R'})],
\end{equation}
and substituting it into Eq. (3) yields 
\begin{eqnarray}
E &=&\frac{N}{2}\sum\limits_{\sigma,\sigma',{\bf k}}
      V({\bf k})\sigma\sigma'Q_{\sigma\sigma'}({\bf k,-k}),
\end{eqnarray}
where $V({\bf k})$ is the Fourier transform of 
$W({\bf R})$:
\begin{equation}
V({\bf k})=\sum\limits_{\bf R}W({\bf R})\exp(i{\bf k\cdot R}) .
\end{equation}

  For a certain ordered structure, only a few $Q_{\sigma\sigma'}({\bf k,-k)}$ 
in Eq. (5) are nonzero under the single-particle approximation.\cite{14,17} 
We assume that this property is still valid in CVM. The assumption is not 
absurd since in experiments there are only a few superlattices for 
a certain ordered structure. Thus Eq. (5) can be simplified. For example, 
for the 1/2\{111\}-type-ordered structure [Fig. 1(a)], all lattice sites 
are divided into two nonequivalent groups. Only the superlattice 
$Q_{\sigma\sigma'}({\bf k}=1/2\{111\})$ are nonzero. Then the energy is 
reduced to
\begin{equation}
E=\frac{N}{2}V({\bf k}=1/2\{111\})\sum\limits_{\sigma,\sigma'}
                   \sigma\sigma'Q_{\sigma\sigma'}({\bf k}=1/2\{111\}).
\end{equation}
By using the symmetry of the ordered structure, we can evaluate 
$Q_{\sigma\sigma'}({\bf k}=1/2\{111\})$ from a small cluster and deal with 
Eq. (7) in CVM. It is noted that $V({\bf k}=1/2\{111\})$ involves any long 
distance interactions. For the 1/3\{111\}-type ordered structure, lattice 
sites are divided into three groups,
\begin{eqnarray}
E & = & \frac{N}{2}V_{1/3}\sum\limits_{\sigma\sigma'}\sigma\sigma'
            \left[ Q_{\sigma\sigma'}({\bf k}=1/3\{111\})
              +Q_{\sigma\sigma'}({\bf k}=2/3\{111\})\right] ,
\end{eqnarray}
where $V_{1/3}\equiv V({\bf k}=1/3\{111\})=V({\bf k}=2/3\{111\})$. 

  Four-point basic clusters (Fig. 1) are chosen in our CVM calculation.

\section{Results}

  The 1/2\{111\}-type ordered structure occurs in both 1:2 and 1:1 
complex perovskites. 
$V_{1/2}\equiv V({\bf k}=1/2\{111\})<0$ is required since it is a necessary 
condition for the 1/2\{111\}-type ordered structure to appear.

  First, the ordering transition type is specified. The 
long-range order parameters (LRO) are defined as
\begin{equation}
\eta=|P_i-P_j| ,
\end{equation}
which represent the occupation-probability difference between nonequivalent 
sites. $\eta$ for B$'$, B$''$, and B$'''$ as functions of the reduced temperature 
$T^*$ are plotted in Fig. 2 when $c=1/3$ and $x=0.25$. (The reduced temperature 
is defined as $T^*=k_BT/|V_{1/2}|$ in this figure and the following Fig. 3.) 
It shows that $\eta$ 
decreases gradually to zero when increasing temperature. So the ordering 
transition is the second order. For 1:1 complex perovskites ($c$=1/2), the 
transition is also the second order.

  Fig. 3 depicts the curves of $\eta$ as functions of the composition 
$x$ when $c$=1/3 and $T^*=0.5$. It shows that $\eta_{\rm B'}$ increases 
when decreasing $x$, while $\eta_{\rm B''}$ and $\eta_{\rm B'''}$ reach their 
maxima near $x=0.25$. The appearance of the maximum in LRO is consistent 
with the experimental findings in BZT-BZ\cite{18} and previous theoretical 
work\cite{10}. For 1:1 complex perovskites ($c=1/2$), $\eta_{\rm B'}$ and 
$\eta_{\rm B''}$ are equal to each other and increase with decreasing $x$, but 
$\eta_{B'''}$ keep equal to zero in all range of $x$. 

  The critical temperatures of the phase transition, $T_c$, can be 
determined at different $x$ values. 
Approximately, $T_c$ decrease linearly with increasing $x$.

  The 1/3\{111\}-type ordered structure is observed only in 1:2 complex 
perovskites. The calculated curves of the long-range order parameters 
$\eta$ as functions of 
the reduced temperature $T^*$ when $x$=0.25 are shown in Fig. 4. (In this 
figure and thereafter, $T^*$ is defined in $T^*=k_BT/|V_{1/3}|$ where $V_{1/3}<0$.) 
It can be seen that LRO of the 1/3\{111\}-type ordered structure suddenly 
drop to zero at the 
transition temperature, which indicates that the transition is the 
first order.

  The 1/2\{111\}- and 1/3\{111\}-type ordered structures are 
separately discussed above. In actual cases, which ordered phase to 
appear is determined by the free energy, { i.e.}, at a 
certain composition, the ordered phase with the minimal free 
energy will become the equilibrium phase and can be observed in 
experiments.

  In the investigations above, the model is parameter-free in the 
sense that $V_{1/2}$ and $V_{1/3}$ define the temperature scale. 
When we consider the 1/2\{111\}- and 1/3\{111\}-type ordered phases 
at the same time, 
a ratio $V_{1/2}/V_{1/3}$ should be introduced. For the purely 
Coulombic interactions\cite{10}, $|V_{1/2}|\approx|V_{1/3}|$; 
for the nearest neighbor interaction\cite{9}, $|V_{1/2}|=2|V_{1/3}|$. 
Adopting $|V_{1/2}|=1.4|V_{1/3}|$, we plot in Fig. 5 the free energy 
curves of the 1/2\{111\}- and 1/3\{111\}-type ordered structures 
for $c=1/3$. It shows that 
the 1/3\{111\}-type ordered phase appears at small $x$ values, and the 
1/2\{111\}-type phase favors larger $x$ values, which agrees with 
results in BMN-BZ\cite{4}. The boundary of $x$ values for the 1/2\{111\}- 
and 1/3\{111\}-type ordered phases is determined by the crossing of 
the curves. The combined phase diagram of the system is shown in 
Fig. 6. The phase region of the 1/3\{111\}-type ordered phase 
near $x=0$ is located inside the region of the 1/2\{111\}-type 
ordered phase.

  When $|V_{1/3}|$ decreases, the phase region of the 1/3\{111\}-type 
ordered structure gets smaller and smaller, and vanishes at last. This 
may be the cases in PMN\cite{8,19,20} and PMT\cite{21} in which 
only the 1/2\{111\}-type ordered phase is observed. Of course, the 
source of the small $|V_{1/3}|$ cannot be solved in the theoretical 
framework of this paper. The effective energy parameters are affected 
by the A-site and O-site ions. Perhaps the different effects of the 
Pb and Ba\cite{22} can provide some clues on this problem.

  For 1:1 complex perovskites ($c=1/2$), the chemical formula is 
favorable to form the 1/2\{111\}-type ordered structure, so 
the situation is different from that of $c=1/3$. The free energy 
curves are depicted in the inserted graphics of Fig. 9
 when $c=1/2$. The curve of the 
1/3\{111\}-type ordered structure wholly lies above that of 
the 1/2\{111\}-type phase. It suggests that the 1/3\{111\}-type 
ordered phase is impossible (or difficult) to form in 1:1 complex 
perovskites such as Pb(Sc$_{1/2}$Ta$_{1/2}$)O$_3$\cite{5} and 
Pb(Mg$_{1/2}$W$_{1/2}$)O$_3$\cite{23}.

\section{Conclusions}

  The 1/2\{111\}- and 1/3\{111\}-type atomic ordering in complex 
perovskites have been studied by the cluster variation method. 
For the 1/2\{111\}-type ordered phase, the transition is the 
second order, and the order parameter reaches its maximum near x=0.25 
for 1:2 complex perovskites. For the 1/3\{111\}-type ordered phase, 
the transition is the first order.The order-disorder 
transition line is found to obey the linear law. The phase diagram 
containing both ordered structures is obtained.

\section*{Acknowledgment}

  This work was supported by the Chinese National Science Foundation. 
We acknowledge Dr. Hui Zheng for many very useful discussions.

\begin{figure}
\caption{Lattice site occupations and basic clusters used in CVM for:  
(a)the 1/2\{111\}-type ordered structure; (b)the 1/3\{111\}-type ordered 
structure. }
\end{figure}

\begin{figure}
\caption{LRO of the 1/2\{111\}-type ordered 
structure as functions of reduced temperature $T^*=k_BT/|V_{1/2}|$ 
at $x=0.25$ and $c=1/3$. }
\end{figure}

\begin{figure}
\caption{LRO of the 1/2\{111\}-type ordered 
structure as functions of $x$ at $T^*=0.5$ and $c=1/3$. }
\end{figure}

\begin{figure}
\caption{LRO of the 1/3\{111\}-type ordered 
structure as functions of reduced temperature at $x=0.25$ and $c=1/3$. 
The reduced temperature is defined as $T^*=k_BT/|V_{1/3}|$.}
\end{figure}

\begin{figure}
\caption{Reduced free energies of the 1/2\{111\}-type (solid line) 
and 1/3\{111\}-type (dashed line) ordered structures as functions 
of $x$ when $c=1/3$. $V_{1/2}=-1.4$, $V_{1/3}=-1.0$, and $k_BT=0.6|V_{1/3}|$. 
Inserted graphics is the case of $c=1/2$. }
\end{figure}

\begin{figure}
\caption{Combined phase diagram when $c=1/3$. $\alpha$, $\beta$, and $\gamma$ 
designate disordered, 1/2\{111\}-type and 1/3\{111\}-type ordered 
phases, respectively. $V_{1/2}=-1.4$, $V_{1/3}=-1.0$. 
The reduced temperature is defined as $T^*=k_BT/|V_{1/3}|$.}
\end{figure}

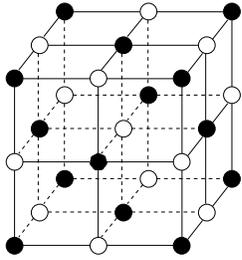
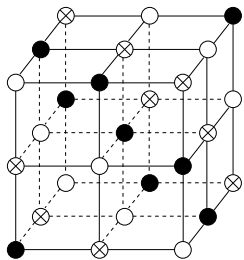
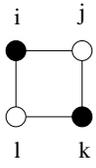
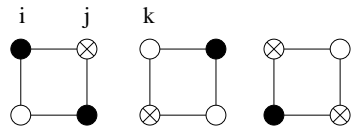
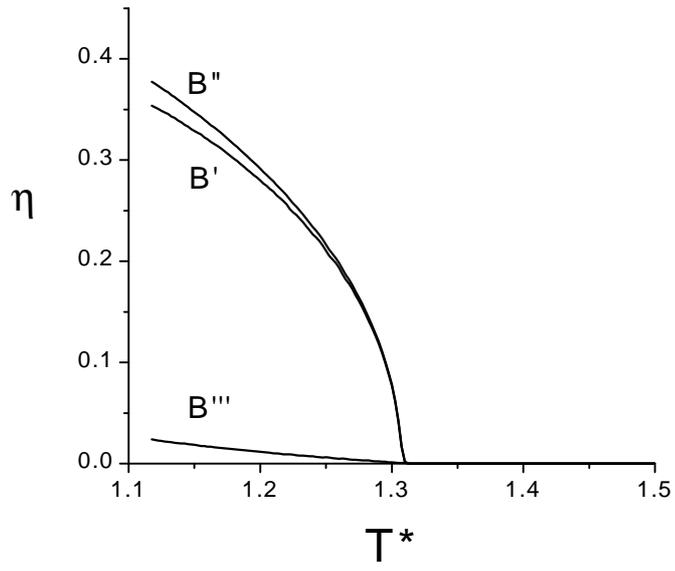
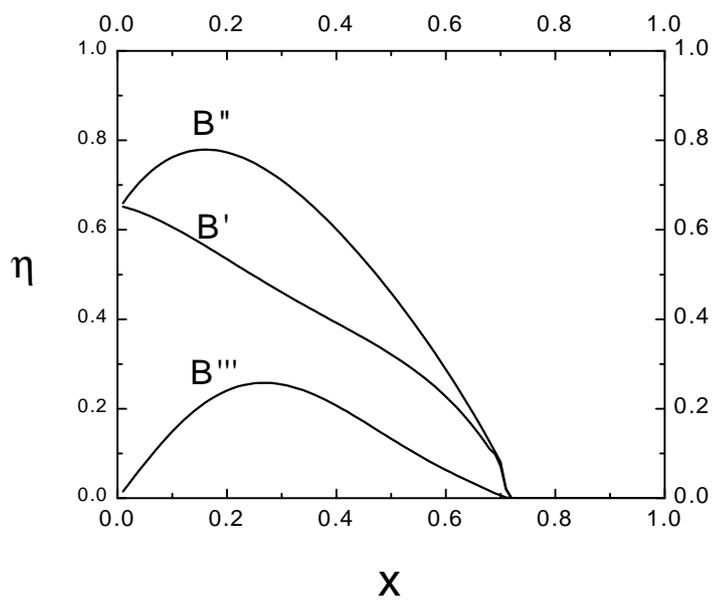
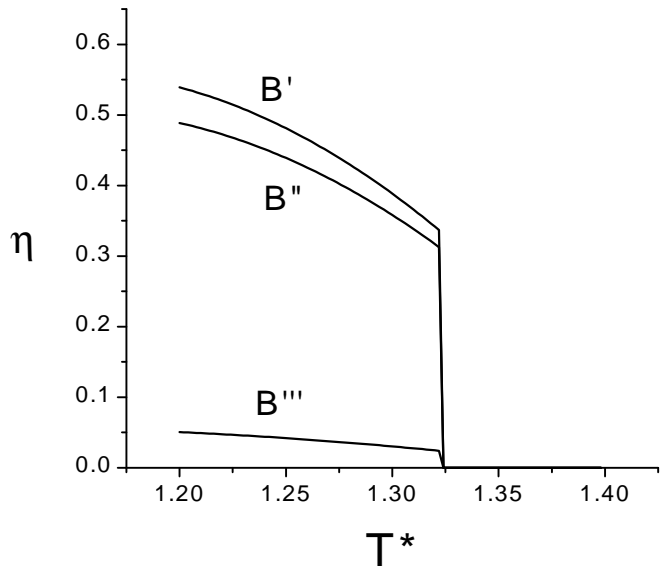
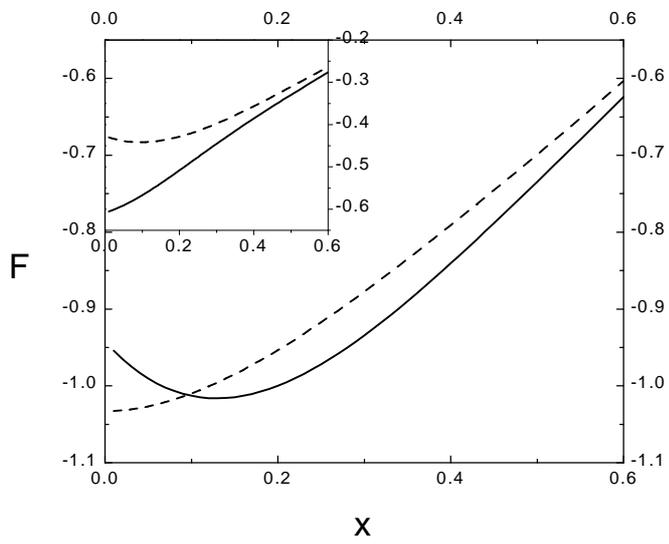
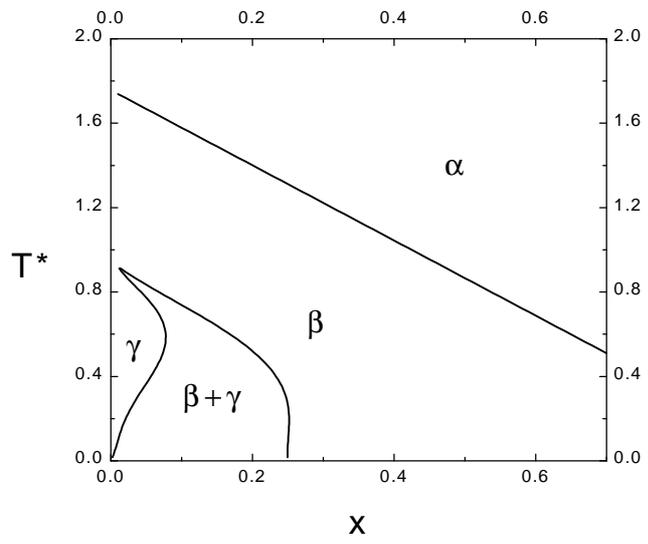
\end{document}